# Diamond-based Detection Systems for Accurate Pulsed X-rays Diagnostics in Radiotherapy

S. Pettinato[a,b], M. Girolami[b], R. Olivieri[c], A. Stravato[d], C. Caruso[d], and S. Salvatori[a,b]

[a] Engineering Dept., Università degli Studi "Niccolò Cusano", 00166 Rome, Italy

[b] Istituto di Struttura della Materia, Consiglio Nazionale delle Ricerche (ISM-CNR), 00010, Montelibretti Rome, Italy

[c] U.O.S.D. Rischio Clinico-Med. Legale-EBM, A.O. "San Giovanni –Addolorata", 00184 Rome, Italy

[d] U.O.C. Radioterapia, A.O. "San Giovanni –Addolorata", 00184 Rome, Italy

Email: marco.girolami@ism.cnr.it, stefano.salvatori@unicusano.it

**Abstract**: The widespread diffusion of precision radiotherapy techniques, geared toward the release of larger dose gradients in shorter time frames, is leading to new challenges in dosimetry. Accurate dose measurements are essential to check for beam anomalies and inaccuracies to ensure treatment efficacy and patient safety during radiotherapy. This work describes the main features of a diamond dosimeter coupled to an extremely compact front-end electronics. The detection system was tested under the X-ray pulses generated by a medical LINAC for both the 6 MV and the 18 MV accelerating voltages. Located in the LINAC's bunker, it eliminates the need for a long cable connection between the detector and the electronics, detrimental for the system response speed. Signal acquisition was performed synchronously with the impinging X-ray pulses with a sampling period as low as 20 µs, allowing for a real-time beam monitoring. The dosimeter demonstrated a very good stability despite the high value of the absorbed dose during the performed experiments (~100 Gy). The measured dose-per-pulse values of 278 µGy and 556 µGy at 6 MV and 18 MV, respectively, are in excellent agreement with the nominal values expected for the LINAC apparatus used for the tests. In addition to single-pulse measurements, fundamental for dynamic radiotherapy, the proposed system also allows for the calculation of both the total collected charge and the photocurrent generated by the detector. In this regards, despite the compactness, it demonstrates its effectiveness as a tool for source diagnostics in terms of both beam intensity and emission timing.

## Contents

## 1. Introduction

Radiation Therapy (RT) is a localized therapy whose purpose is to cause the necrosis of tumor cells through the use of ionizing radiation, while minimizing the dose delivered to surrounding healthy tissues. External beam RT is based on linear accelerators (LINACs) in which electron packets are generated and accelerated up to several megavolts. X-ray pulses are then produced by the Bremsstrahlung effect by the collision of accelerated electrons on a heavy metal target. In medical LINACs, the produced high-energy X-ray beam is shaped with a multi-leaf collimator. Following a particular conformation and treatment duration, the customized beam is directed onto the patient's tumor mass. In recent decades, sophisticated RT techniques have been developed to achieve the necessary precision for cancer treatments. Currently, Intensity Modulated RT (IMRT) [1], Volumetric Modulated Arc Therapy (VMAT) [2], and Stereotactic Body RT (SBRT) [3], provide the precision required in modern radiotherapy medicine. Treatments are achieved with very narrow beams and special multi-laminar collimators, allowing for highly conformal dose distributions that can target also complex geometries of solid tumors. Such advanced techniques give rise to new challenges, especially in the area of Quality Assurance (QA) for the definition of patient-specific RT treatment plans [4]. Indeed, both the high-conformity and the high-dose of RT-beams lead to possible uncertainties in the different steps of a treatment. Therefore, in order to precisely define the radiation dose that will be delivered to the patient, the use of accurate detection systems becomes crucial for the definition of the treatment plan.

The photon beam generated by a medical LINAC is composed of a series of pulses. Each pulse has a duration of a few microseconds with a repetition rate in the range $60 - 1000$ Hz. Especially for dynamic treatments, the accurate measurement of the dose delivered by each single pulse of photons is fundamental. Dose measurements in RT are commonly performed by means of ionization chambers [5, 6], silicon photodiodes [7], and radiochromic films [8]. The main drawback of the latter is that they are off-line dosimeters, therefore not enabling a real-time beam monitoring, whereas in silicon-based devices, due to the lack of tissue-equivalence, sensitivity is energy-dependent. Ionization chambers offer the best accuracy, and represent the gold standard devices employed for dosimetry in RT. However, even the smallest ionization chambers have a volume of several mm$^3$ and require a bias voltage greater than 100 V [9]. Moreover, despite their fast collection times (in the range $10 - 100$ μs) [10], they are commonly coupled to relatively slow high-precision electrometers with a response time of tenths of a second [11].

It appears then clear that, to perform pulse-by-pulse dose measurements, and thus identify possible beam anomalies even in short time intervals, fast dosimeters coupled to an adequate front-end electronics are then essential. In this regard, diamond represents the most suitable material for the realization of high-speed detectors [12] as it shows a response time in the nanosecond range [13, 14]. Moreover, the development of the Chemical Vapor Deposition (CVD) technique for diamond synthesis allows to obtain very high-quality diamond samples, all with specific controlled and reproducible properties at a relatively low cost in comparison to natural specimens. Very significantly, the concentration of impurities in high-quality CVD-diamond (less than 5 ppb and 0.5 ppb for nitrogen and boron, respectively, as declared for electronic- or optical-grade samples [15]) is three orders of magnitude lower than that of the best natural diamond (type-IIa). CVD-diamond dosimeters represent a mature technology in the field of RT dose measurements [16-18], for their special characteristics such as tissue equivalence, high dose response, high spatial resolution, high radiation hardness (up to 10 MGy), and high cohesion energy ($\approx$ 43 eV). CVD-diamond also demonstrates its maturity for the fabrication of X-ray [19-21], charged particle [22-25], and UV detectors [26-27].

In the literature, solutions in RT for pulse-by-pulse detection based on scintillators and fiber-optic sensors have been reported [28-31]. However, relatively complex optical detection systems and Cherenkov-radiation removal techniques are required. Velthuis et al. [32] proposed a CVD-diamond based measurement system for pulse-by-pulse dosimetry, with a response time of the order of tenths of ms. In previous works [33-34], we described preliminary characterizations, performed under 6 MV X-ray photons, of a CVD-diamond dosimeter coupled to a first version of the front-end and readout electronics introduced in this paper. Here, an in-depth systematic investigation was carried out on the proposed prototypal detection system by evaluating the performance under either 6 MV or 18 MV X-ray pulses sourced by a medical LINAC, at different dose rates (DRs). This paper aims at illustrating the flexibility of our system, able to provide a real-time dose-per-pulse measurement, which is crucial for dynamic radiotherapy treatments, as well as to quantify both the dose-rate and the total dose delivered during the treatment. In addition, this work demonstrates that the system is an effective tool for pulsed X-ray diagnostics, allowing for a real-time evaluation of the beam source characteristics in terms of both emission timing and pulse.

## 2. Experimental set-up

A Clinac iX (Varian Inc.) installed at the Radiotherapy Department of "San Giovanni – Addolorata" Hospital in Rome (Italy) was used to verify the performance of the CVD-diamond based detection system for X-ray pulses described in this work. The LINAC was set at either 6 MV or 18 MV electron acceleration voltage with DRs in the $1 - 6$ Gy/min range. The diamond detector was placed in a Plexiglas® phantom and positioned at the LINAC isocenter (at a distance of 100 cm from the X-ray source) under a 10×10 cm$^2$ field. The impinging photon beam direction was perpendicular to the diamond plate surface.

### A. Detector and front-end readout electronics details

The X-ray detector is based on an optical grade single-crystal CVD-diamond sample (Element Six). 300 nm thick Ag contacts were fabricated on the top and bottom surface of the 4×4×0.5 mm$^3$ sample by means of sputtering deposition technique. A shadow mask was used to define the circular contact geometry with 3.2 mm in diameter. The resulting active volume of the detector is around 4 mm$^3$. Two thin wires soldered to the outer shield and the inner conductor of a 3.6 m long triaxial cable were glued to device contacts with silver paste. To assure mechanical stability, before encapsulating the detector into a PMMA cylinder (∅ 9 mm, 5 cm long), the two wires were glued with epoxy resin (Epotek® 301). Finally, the same epoxy resin was used to fill the free volume of the cylindrical housing.

**Figure 1** illustrates the schematic of the implemented prototypal setup. Diamond dosimeter and readout electronics were placed inside the bunker where the Clinac iX is installed. A 20 m long LAN cable was used for remote interfacing with a PC located outside the bunker, in a room where the LINAC console is installed. The front-end, based on the precision integrator IVC102 chip, implements the idea described elsewhere [35-36]: exploiting the sync signal generated by the LINAC, integration is performed only in a period around each pulse. By strongly reducing the noise contribution, the employed method ensures better measurement accuracy, as well as single-pulse diagnostics.

As shown in **Figure 1**, the outer shield of the triaxial connector is used for detector biasing. A fixed value of 10 V was used for the measurements presented in this work and it was obtained by means of an on-board REF102 precision reference voltage device. The inner conductor of the triaxial cable is connected to the integrator input, whereas the inner-shield is at ground potential, in order to minimize leakage currents at the front-end input. For a bias voltage $V_{bias} > 0$, integrator output generates negative ramps. Then, an inverting amplifier based on the fast LT1995 precision gain-selectable amplifier is inserted between the IVC102 output and the analog-to-digital converter (ADC) input. The insertion of the LT1995 allows to attenuate or amplify the integrator output voltage simply by changing the connections to the chip; therefore, for future developments, the system can be easily adapted to the signal level of the detector according to the particular operating condition. Synchronized with the LINAC sync signal, a specifically programmed timer generates *S1* and *S2* signals used to define both the integration time and the holding period [35]. The ADC and the timing control circuitry are integrated into an LPC845 microcontroller also used for data processing and transfer.

A preliminary in-the-lab characterization of the prototype was performed to evaluate the integrating capacitance value $C_{INT}$, by injecting a known constant current sourced by a Keithley 6221A. To eliminate any offset contribution, the output voltage $V_O$ was measured for both positive and negative input currents. By calculating the slopes of the generated voltage ramps, a $C_{INT} = (88.49 ± 0.01)$ pF was estimated (in very good agreement with the 90 pF nominal value obtained by parallel connection of the IVC102 internal capacitors, $C_2$ and $C_3$, used for our experiments). As a further development, an on-board precision current source [37] will be integrated in the system, thus allowing for an automatic periodic calibration procedure of the front-end electronics.

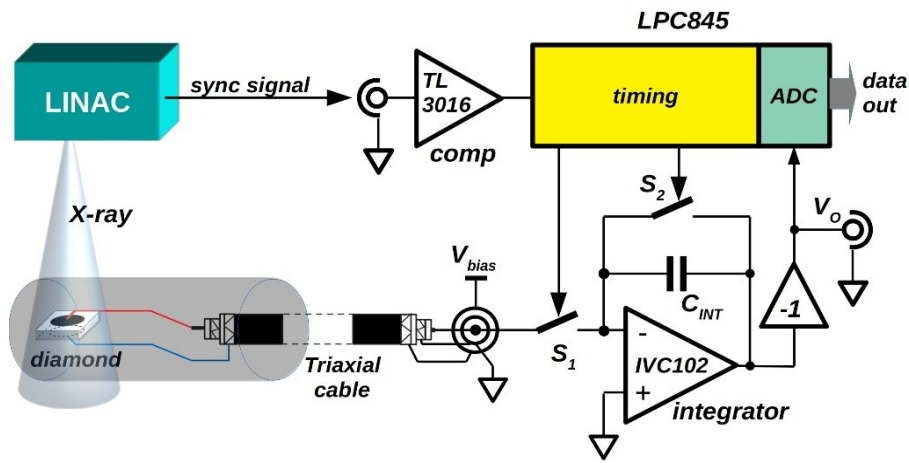

**Figure 1.** Block diagram of the front-end readout electronics used for gated-integration of the
collected charges in the diamond dosimeter.

### B. Gated-integration measurements

Plots reported in **Figure 2** show an example of the integrator output voltage $V_O$ as a function of time acquired by means of a DSO-X-3024A digital oscilloscope (Keysight) with the dosimeter irradiated by either a 6 MV (blue curve) or a 18 MV (red curve) single X-ray pulse. As expected [35], the integrator output voltage starts to increase about $12 - 13$ µs after the sync signal rising-edge (green pulse). It is important to observe that VO flattens out after a further 10 µs, whereas X-ray pulses are expected to be only 4 µs wide [35]. In order to get a deeper insight into parameters affecting integrator response, the circuit was simulated using SPICE. The model included an op-amp with a gain-bandwidth product and a slew rate of 2 MHz and 3 V/µs, respectively, as indicated for the IVC102 chip [38]. The detector was modeled as an ideal current pulse generator (4 µs in duration and 100 ns for both the rise and the fall time) in parallel with a capacitor and a resistor. For the latter, a value of $10^{13}$ Ω was used, as inferred by the slope of the voltage-current linear characteristics obtained for the diamond detector in dark conditions up to 100 V. Dashed lines of **Figure 2** refer to results obtained including an input series resistance of 1.7 kΩ and a sensor capacitance of 350 pF. The former fits the on-resistance of IVC102's S1 MOS switch, whereas the latter mainly reflects the cable capacitance, estimated to be about 100 pF/m, in good agreement with what reported for triaxial cables [39]. Dot-dashed line illustrates the simulated result for a 20 m long cable (which introduces a total capacitance of about 2 nF). The effect underlines the importance of a relatively short connection between the detector and the readout electronics for single pulse diagnostics within a few microseconds, as adopted in this work.

By the voltage values $V_{O(HOLD)}$ acquired during the hold periods (t ≥ 32 µs), for an integration capacitance $C_{INT}$ = 88.49 pF, collected charge values of 84.07 pC and 163.71 pC were evaluated for photons of 6 MV and 18 MV, respectively. Therefore, charge-per-pulse under 18 MV X-ray pulse is substantially twice the one collected at 6 MV. No charge injection jumps are observed when S1 switch is opened (hold period), mostly due to the high value of the capacitance seen by the integrator input. Very significantly, a negligible voltage drop of about 1 mV/s, mainly affected by the op-amp input bias and the detector bias current, was evaluated for data recorded up to t = 0.5 ms.

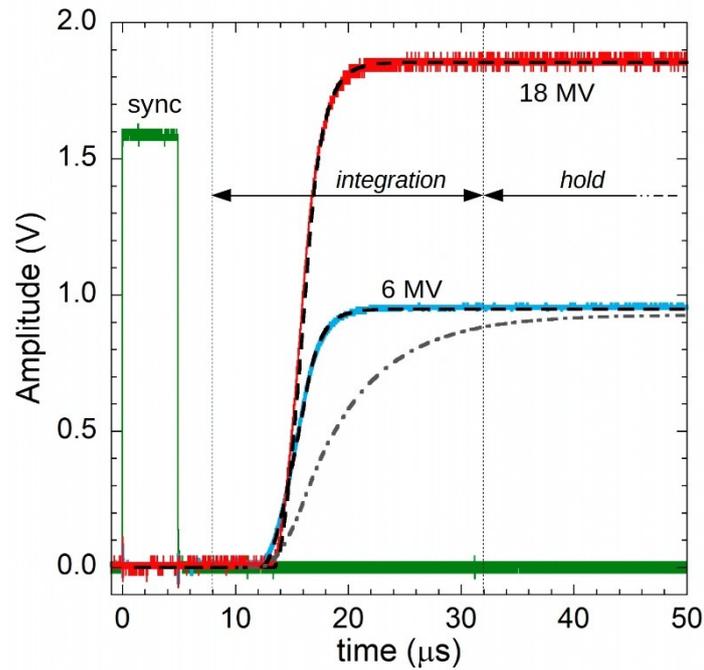

**Figure 2.** Typical output voltage $V_O$ of the proposed readout electronics for 6 MV (blue) and 18 MV (red) X-ray pulses impinging on the CVD-diamond dosimeter. Sync signal (green) is used to synchronize the *S1* and *S2* control signals. Dotted lines represent simulations (see text).

As previously mentioned, the Clinac iX apparatus generates X-ray pulses synchronized with the *sync* signal available at the LINAC console. **Figure 3** illustrates the sync (red peaks) and $V_O$ (blue peaks) signals captured at the two electron accelerating-voltages of 6 MV and 18 MV, at all the available DRs ($1 - 6$ Gy/min) on a hundred acquisitions of 1 s each, visually summarized in a few tens of ms in **Figure 3** for clarity. Characterization was carried out selecting an oscilloscope time base of 1 s. The LINAC guarantees the DR amount by periodic suppression of pulses, as highlighted by $V_O$ peak-positions. As can be seen, excellent periodicity was observed for all the acquisitions performed during the characterization; indeed, we evaluated an error approximately equal to 1 μs for the $V_O$ signal period, to be attributed exclusively to the oscilloscope limited resolution in the chosen time base.

It is worth observing that the LINAC produces pulses having a periodicity only dependent on the DR, with a pulse-repetition-rate (PRR) for 18 MV photons which is exactly half the one used for 6 MV photons. Such a behavior completely agrees with a detector's charge collection, hence an absorbed dose, in 2:1 ratio at the cited photon energies, as underlined in the examples reported in **Figures 2** and **3**. The number *n* of generated pulses-per-second can be written as:

$$n = k \, f_{sync} \, DR \qquad (1)$$

where $f_{sync}$ is equal to 360 Hz and 180 Hz for 6 MV and 18 MV, respectively and $k$ = 1/6 min Gy$^{-1}$ Hz$^{-1}$ for dose rate DR reported in Gy/min. By Eq. 1, the nominal dose-per-pulse value, $D_P$, can be calculated as 0.1/$f_{sync}$, resulting in 277.78 μGy (for 6 MV photons) and 555.56 μGy (for 18 MV photons).

For the Clinac iX apparatus, it is also worth stressing here that, also at the highest dose-rate, i.e. the highest PRR, a time interval of more than 2.7 ms can be used for signal acquisition, processing, and data

- 7 -

transfer, thus enabling a real-time diagnostics of impinging X-ray pulses for the proposed detection system, as hereafter described.

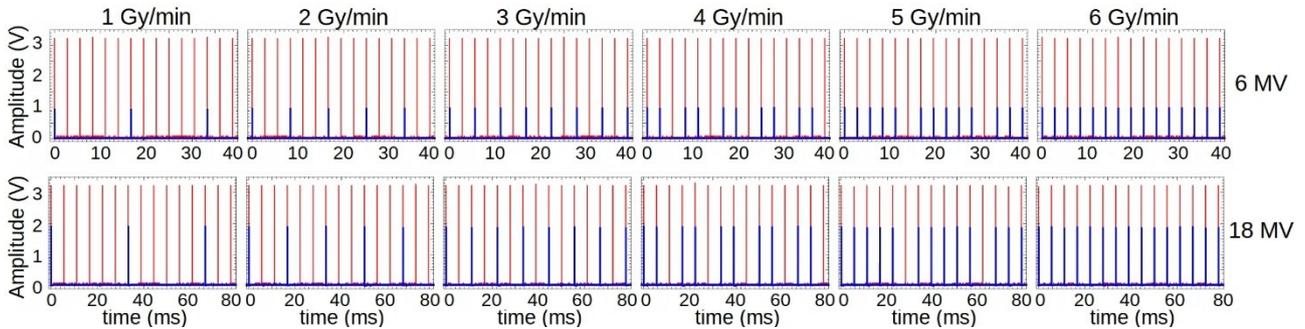

**Figure 3.** *Sync* pulses (red) and $V_O$ output signal (blue) at different dose-rates for X-ray photons emitted at 6 MV and 18 MV.

## 3. Results and discussion

The proposed detection system was also used to monitor the generated photocurrent and the total charge collected during irradiation. As stated before, the charge collected at each impinging X-ray pulse is given by $C_{INT} \times V_{O(HOLD)}$, where $C_{INT}$ = 88.49 pF. The current is calculated as the charge-per-pulse value divided by the elapsed time from the previous pulse. **Figure 4** illustrates the detector behavior in the very first minutes of X-ray irradiation for a total absorbed dose of 2 Gy. In particular, for 6 MV photons, 5.046 nA and 10.081 nA average values were recorded at 1 Gy/min and 2 Gy/min dose-rates, respectively. The observed 0.45 nA peak-to-peak photocurrent amplitude has not to be attributed to noise, but it reflects the real X-ray pulse amplitude change. Under 18 MV photons, in comparison to the previous ones, photocurrent amplitudes acquired at the same dose-rates are about a 5% lower, and equal to 4.805 nA and 9.618 nA, with a peak-to-peak amplitude around 0.24 nA. Such a result is tentatively attributed to a decrease of detector efficiency for high-energy photons, due the extremely low strength of the electric field applied to the detector. The collected charge related to the total dose (see the plot on the bottom of **Figure 4**) shows an excellent linearity ($R^2 \geq 0.99995$ for the four plots) indicating an optimal stabilization of the delivered dose performed by the LINAC apparatus. It is worth noting that the detector photoresponse does not show any dependence on time, indicating a very good stability, as remarked in the following section.

The results shown in **Figures 2** and **4** underline the versatility of the proposed prototype. The system allows both pulse-by-pulse measurements and evaluation of the total charge, as well as of the instantaneous current of the detector, with a time resolution limited only by the pulse repetition rate of impinging X-rays. In this work, further investigations on the detection system features were carried out. In particular, characterizations were made according to four different steps:

A) Irradiation of the dosimeter by 6 MV photons, for DRs in the 1 – 6 Gy/min range, with steps of 1 Gy/min. At each DR, the diamond detector was irradiated for about 90 s.

B) The same procedure as A) was repeated under 18 MV photons.

C) For comparison, the diamond detector was then connected to a Keithley 6517A electrometer and biased at 10 V. At a fixed DR = 3 Gy/min, the detector collected charge was measured for dose values from 1 to 6 Gy, with steps of 1 Gy, under 6 MV X-rays.

D) The same procedure as C) was repeated under 18 MV photons.

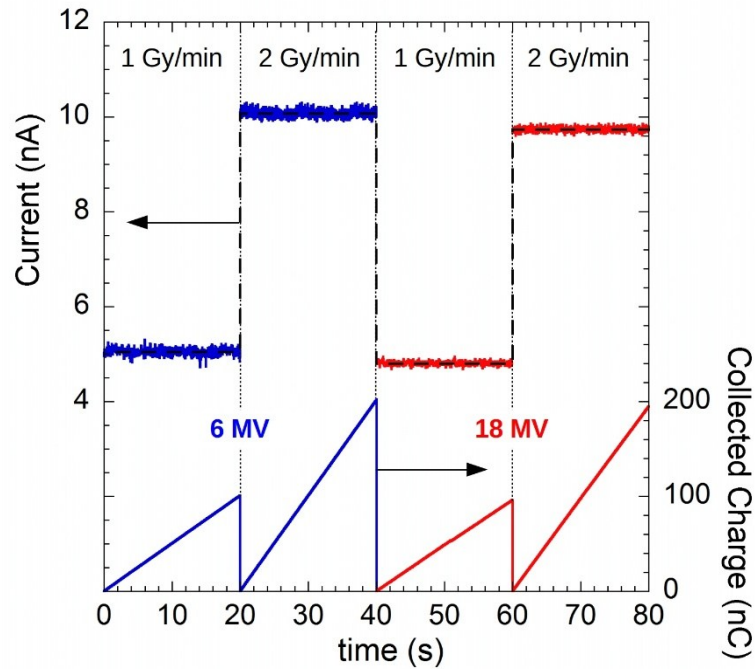

**Figure 4.** Detector photocurrent and collected charge at different dose rates for X-ray photons emitted at 6 MV and 18 MV as processed by the proposed detection system.

During the phase A), at each DR, consecutive pulses within the first 5 s were acquired soon after the LINAC's beam-on command. However, to evaluate its response stability, the detector was further irradiated for about additional 1.5 minutes at each DR. To have a number of pulses acquired comparable to that obtained at 6 MV, during irradiation under 18 MV photons (phase B) the acquisition time was doubled to 10 s at each DR. It is worth noting that, at each DR, being irradiation time about 90 s, the dosimeter absorbed a significantly high dose during these tests. More specifically, at the end of both phase A) and phase B) the absorbed dose was around 30 Gy. During the last C) and D) steps, being DR = 3 Gy/min, the dose absorbed by the detector at each phase was 21 Gy. Therefore, a total dose greater than 100 Gy was absorbed by the diamond during the performed measurement session.

### A. Detector stability

Diamond dosimeters usually require a priming process, i.e. a pre-irradiation period under high-energy photons or particles, to stabilize their sensitivity [40]. Based on the acquisitions performed during the aforementioned phases, a very good response stability of the detector was found, as highlighted by data summarized in **Figure 5**. Graphs indicated by odd and even numbers refer to pulses' height acquired at the beginning and at the end of a phase, respectively. Data recorded by means of the proposed detection system (1-4) refer to the processing of the first and the last 200 pulses at 6 MV and 18 MV X-ray photons. Conversely, for the measurements performed with the electrometer (phases C-D) the sampling period was programmed to be $\Delta t$ = 0.5 s. In such a case, taking into account the periodicity of signals generated by Clinac iX (see **Figure 3**), the reported charge-per-pulse values (5-9) were calculated by dividing the measured $Q(t+\Delta t) - Q(t)$ charge increase by the number $N_{0.5}$ of pulses generated by the LINAC in the same period $\Delta t$. During these phases, being DR = 3 Gy/min, $N_{0.5}$ was equal to 90 and 45 for 6 MV and 18 MV photons, respectively. In **Figure 5**, the dose absorbed by the detector during each phase is also reported. As evidenced by the two horizontal dashed lines, the average values of the charge-per-pulse collected by the detector remained constant despite the ~100 Gy of total absorbed dose, indicating a very good stability of the diamond detector.

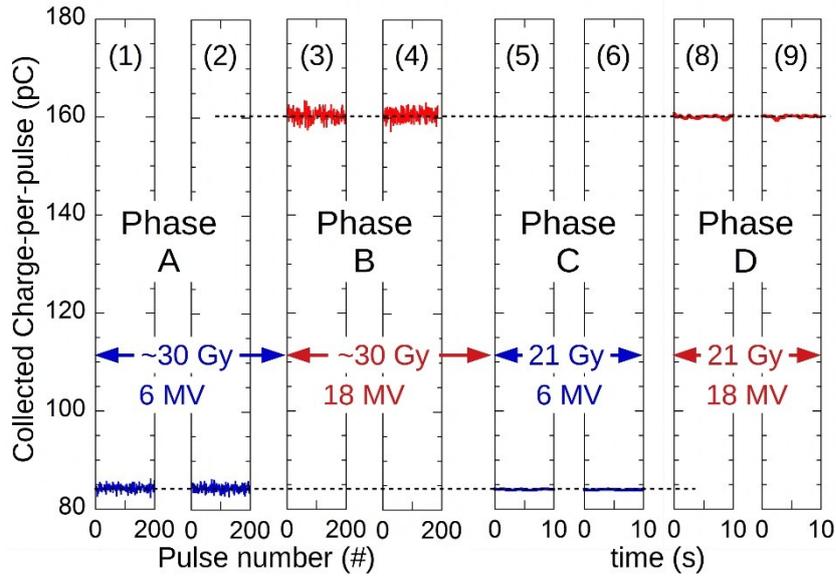

**Figure 5.** Collected charge-per-pulse values at 6 MV (red) and 18 MV (blue) X-rays recoded by the proposed prototype (1-4) and those estimated by measurements carried out by means of a Keithley 6517A electrometer (5-9) (see text). Odd and even numbers refer to the beginning and the end, respectively, of each phase (from A to D). The total detector's absorbed dose in each phase is also reported.

### B. Detector sensitivity

The detector sensitivity to LINAC X-rays was also evaluated. Data reported in **Figure 6** highlight an excellent linearity of the collected charge, $Q$, per unit dose at the fixed DR of 3 Gy/min. In the 1 – 6 Gy range, data refer to the charge measured by means of the 6517A electrometer. Conversely, for a dose lower than 1 Gy, if $q_i$ is the charge measured for the $i^{th}$ pulse, the quantity

$$Q(D_m) = \sum_{i=1}^{m} q_i \qquad (2)$$

represents the accumulated charge measured by our system for a given number, $m$, of impinging pulses, with an expected dose provided by the LINAC given by $D_m = m \times D_P$. Note that the two values indicated with arrows refer to the dose released by a single pulse. The diamond dosimeter sensitivity was evaluated to be (75.16 ± 0.02) μC Gy$^{-1}$ cm$^{-3}$ for 6 MV X-ray photons, and (71.64 ± 0.01) μC Gy$^{-1}$ cm$^{-3}$ for 18 MV photons, respectively. It is worth recalling here that the difference between the estimated slopes of data acquired by the proposed prototype and the Keithley 6517A is lower than 0.2%, highlighting an excellent agreement between the two instruments.

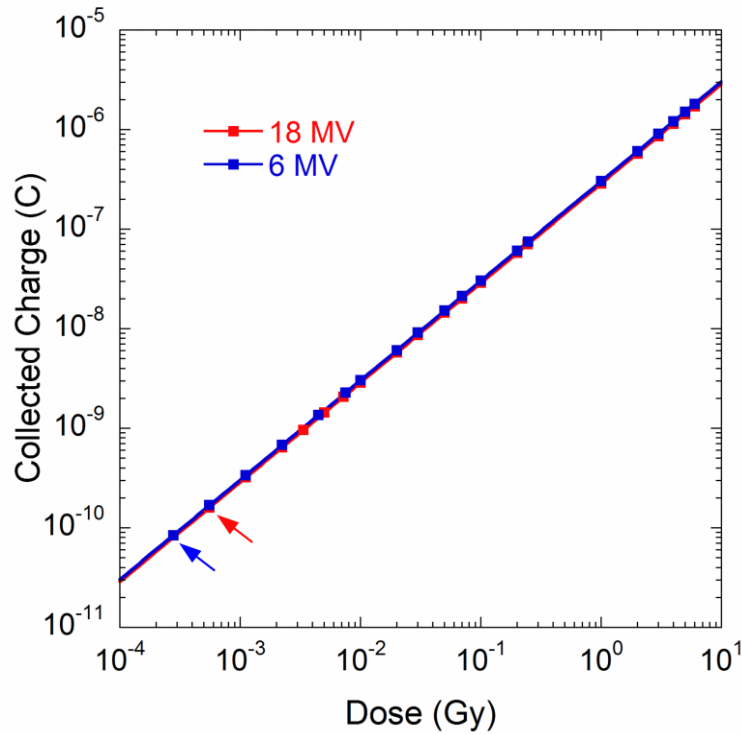

**Figure 6.** Collected charge as a function of dose for X-ray photons emitted at 6 MV and 18 MV. Data for dose lower than 1 Gy have been acquired by the proposed prototype, whereas a Keithley 6517A electrometer was used in the 1-6 Gy range.

### C. Dose-per-pulse measurements

The excellent sensitivity of the diamond detector allowed us to evaluate with our prototypal detection system the dose delivered by each single pulse. Data recorded at all possible DRs delivered by the LINAC are reported in **Figure 7**, from the first acquisition at the maximum DR (6 Gy/min) to the last acquisition performed at the lowest DR (1 Gy/min). Vertical dashed lines indicate the change of DR values during the tests. In the figure, continuous lines represent the cumulative sum of the dose-per-pulse. The ~1.8 Gy and ~3.5 Gy dose values, at 6 MeV and 18 MeV energies, respectively, are consistent with the maximum dose ranges typically delivered in a single radiotherapy treatment session. Recorded data highlight the ability of the LINAC apparatus to adjust the dose in fractions of a second, thus maintaining the delivered dose on a stable average value. In particular, regardless the DR, as observed by the slope of cumulative sum – as well as by the calculated mean values (horizontal dashed white lines) – doses of about 0.278 mGy/pulse and 0.556 mGy/pulse are found at 6 and 18 MeV, respectively. Such results are in excellent agreement with the expected DP nominal values mentioned at the end of Section 2.

Previous results highlight the effectiveness of the proposed system to monitor the intensity of single pulse. For data collected during these tests, it is interesting to observe that at the beginning of some irradiation sessions the apparatus delivered dose-per-pulse values slightly lower than the mean amplitude DP. This effect appeared more pronounced the lowest DRs, as indicated, for example, by the gray circle in **Figure 7** and reported as the detail of 600 pulses in **Figure 8**. The cumulative mean value calculated in this frame is shown by the red curve. Although this appeared as the worst case observed in the entire measurement session, it is noteworthy that the LINAC was able to settle the average dose value to 99% of the nominal set point in less than 40 pulses, i.e. 0.3 s, and to the 99.5% in less than 0.8 s. Remarkable is also the ability of the prototype system to provide accurate and real-time information on the beam intensity stability, a fundamental feature in the use of LINACs for dynamic radiotherapy treatments.

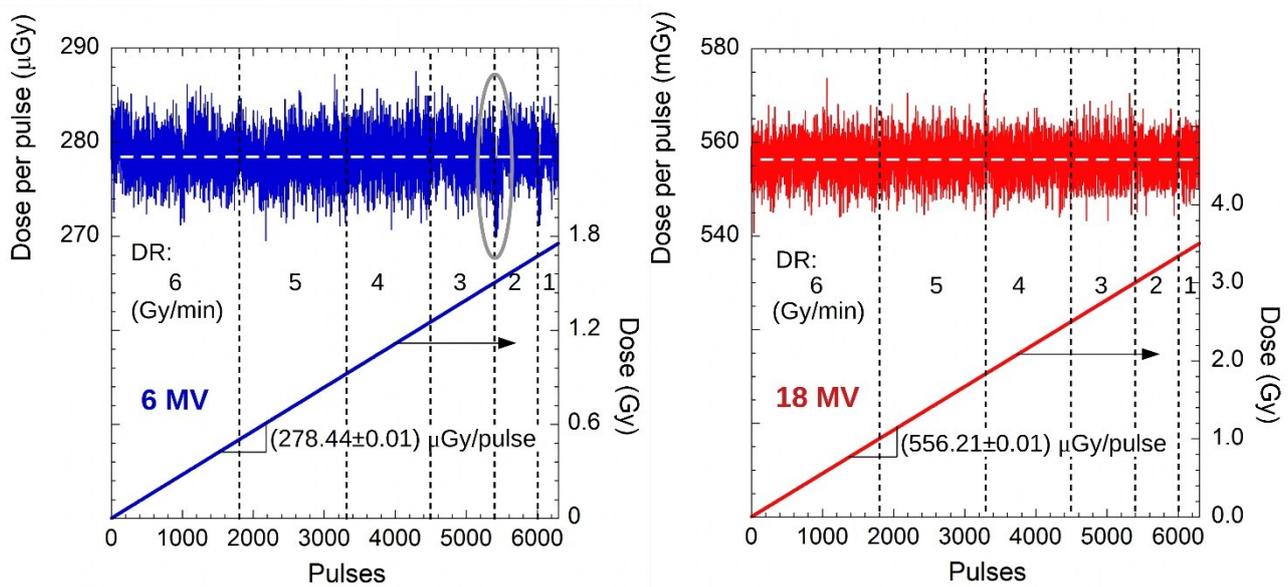

**Figure 7.** Dose per pulse and total collected dose recorded by the proposed prototypal system for different DRs. An average value of about 0.278 and 0.556 mGy/pulse, at 6 MV and 18 MV, respectively, is found, Cumulative sum of dose (continuous lines) displays a slope in excellent agreement to the expected DP value.

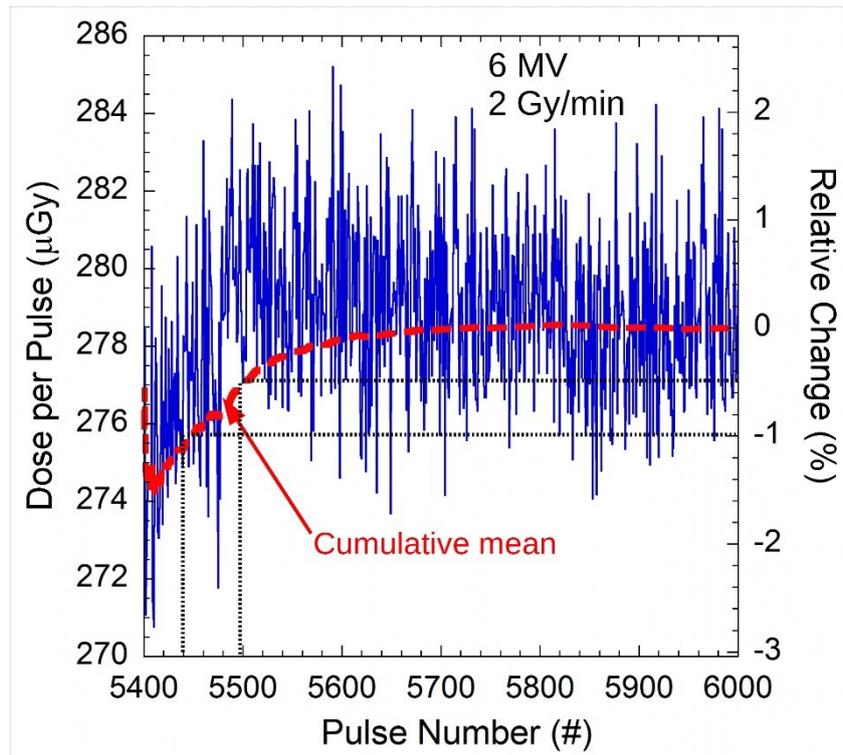

**Figure 8.** Detail of the dose released at the beginning of 6 MV irradiation at 1 Gy/min (see gray circle in **Figure 7**). The cumulative mean (in red) highlights the LINAC's ability to maintain the average dose at the desired value within a few tens of pulses.

## 4. Conclusions and further developments

A measurement system based on a single crystal CVD-diamond dosimeter and a dedicated front-end electronics was developed and fully characterized on the field under X-rays produced by a medical LINAC (Varian, Clinac iX) used for radiation therapy. The main advantage, resulting from the simplicity of the proposed electronics, is the extremely low number and low cost of the components used. This guarantees, in addition to compactness and cost effectiveness, both greater reliability and greater measurement accuracy. In this regard, experimental results demonstrate that the system allows for accurate real-time pulse-by-pulse dose measurements, thus meeting the demands of QA in dosimetry, particularly stringent in modern dynamic RT. Indeed, due to the availability of the *sync* signal at the LINAC console, the implemented synchronous measurement method allows the charge collected by the detector to be integrated and to be processed before the next pulse arrives. (Work is also in progress to implement an auto-sync solution for the measurement in case a synchronization signal is not available from the LINAC apparatus). Simulations demonstrate that only the length of the detector-to-electronics connection degrades the system response time. In our case, the prototype, placed right in the bunker where the X-ray source is installed, allows the monitoring of the beam also for a pulse-repetition-rate well above 1 kHz.

It is worth highlighting the very good response stability of the diamond detector irradiated by about 100 Gy. By a proper calibration procedure of both the detector and the integrating capacitance, dose-per-pulse values of 0.278 mGy and 0.556 mGy were evaluated at 6 MeV and 18 MeV energies, respectively, in excellent agreement with the expected nominal values of the LINAC. The system has also proven to be effective in measuring any changes in beam intensity of even a few fractions of a percent, as occurs in the case of a soft-start-mode implemented in the medical LINAC used for tests. Thus, the proposed instrument is effective for X-beam diagnostics in terms of both intensity and emission time. Moreover, the detection system was successfully tested for the measurement of the cumulative collected charge and the instantaneous photocurrent as well.

Finally, it's worth recalling here that CVD diamond, also with 3D contact structures, has been shown to be effective for the detection of high-energy electrons [25, 41]. In this regard, by exploiting the same source used in this work, further tests of the proposed detection system are under development. Finally, work is in progress to further validate the proposed front-end electronics to process the signal generated by commercial dosimeters, aimed at overcoming the speed limitations of standard electrometers [11].

## Acknowledgment

The authors would like to acknowledge Fabrizio Imperiali for the skilful assistance in setting the LINAC setup to the experimental needs.